\documentclass[twocolumn,showpacs,preprintnumbers,amsmath,amssymb]{revtex4}
\usepackage{graphicx}% Include figure files
\usepackage{dcolumn}% Align table columns on decimal point
\usepackage{bm}% bold math

\begin{document}

\preprint{APL}

\title{Spin injection in a single metallic nanoparticle: a step towards nanospintronics}% Force line breaks with \\

%\author{Ann  Author}
%\affiliation{%

%\author{Charlie Author} \homepage{http://www.Second.institution.edu/~Charlie.Author}\affiliation{

\author{A. Bernand-Mantel}

\author{P. Seneor}
%\email{pierre.seneor@thalesgroup.com}
\author{N. Lidgi}

\author{M. Mu\~{n}oz}
\altaffiliation[Now at ]{Laboratorio de F\'{i}sica de Sistemas
Peque\~{n}os y Nanotecnolog\'{i}a, Consejo Superior de
Investigaciones Cient\'{i}ficas, Serrano 144, 28006 Madrid, Spain}
\author{V. Cros}

\author{S. Fusil}
\altaffiliation[Also at ]{Universit\'{e} d'Evry, Bat. des Sciences,
rue du pere Jarlan, 91025 Evry, France}

\author{K. Bouzehouane}

\author{C. Deranlot}

\author{A. Vaures}

\author{F. Petroff}

\author{A. Fert}
 \affiliation{Unit\'{e} Mixte de Physique CNRS/Thales and Universit\'{e} Paris-Sud 11\\ 91767, Palaiseau, France}

\date{\today}% It is always \today, today,

\begin{abstract}
We have fabricated nanometer sized magnetic tunnel junctions using a
new nanoindentation technique in order to study the transport
properties of a single metallic nanoparticle. Coulomb blockade
effects show clear evidence for single electron tunneling through a
single 2.5~nm Au cluster. The observed magnetoresistance is the
signature of spin conservation during the transport process through
a \emph{non magnetic} cluster.
\end{abstract}

\maketitle

%\section{\label{sec:level1}}
Spintronics debuted with the discovery of giant
magnetoresistance\cite{Baibich88} effect in magnetic multilayers in
which a single dimension was reduced to the nanometer range. This
field was then extended to structures with two reduced dimensions
like nanowires\cite{Piraux94} and
nanopillars\cite{Katine00,Grollier01} or
nanotubes\cite{Tsukagoshi99}. Today, a challenge for spintronics is
the study of spin transport properties in structures based on 0D
elements in which the three dimensions have been reduced. In
particular, we have in mind systems in which the reduction of the
size leads to both Coulomb blockade and spin accumulation
effects\cite{Barnas98,Imamura99,Brataas99}. Transport studies on
systems including mesoscopic islands\cite{Zaffalon05,Ono97,Chen02_3}
or granular films\cite{Mitani98,Yakushiji05,Schelp97} have paved the
way to understanding the effect of confinement on charge and spin
transport properties in metallic nano-objects. However, so far, very
few techniques allow to contact a single isolated nanometer sized
object\cite{Ralph95,Desmicht98,Delft01} to study the effect of
confinement on spin transport.

\begin{figure}
  \includegraphics[width=8cm,keepaspectratio=true]{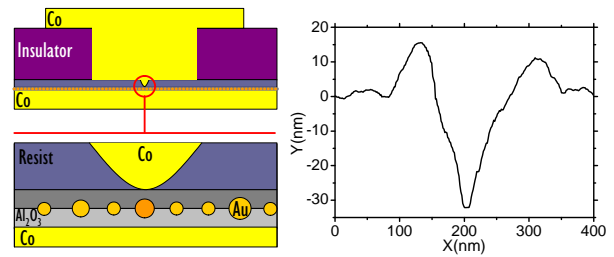}
\caption{Left: Schematic cross-section of the whole patterned
structure showing the top/bottom Co electrodes and the 2D assembly
of nanoparticles embedded in a thin alumina layer. The circle
represents the zone zoomed in the bottom drawing. Right: Cross
section of an AFM tapping mode scan of a nanoindent before filling
with the Co top electrode. The effective nanocontact cross section
is less than 10nm.}
  \label{fig:dessin-structure}
\end{figure}

In this letter we present the experimental achievement of a new
technique allowing us to inject and detect spins in a single
isolated nanometer sized cluster. We then obtain information on both
spin and single electron transport in the nanoparticle. In this
technique, a ferromagnetic nanocontact with a cross section of $\sim
5-10$~nm in diameter is created on a bilayer associating a cobalt
layer and an ultrathin alumina layer in which a 2D assembly of gold
nanoparticles is embedded (see Fig~\ref{fig:dessin-structure} for a
sketch). As we will show, this structure allows the tunneling of
electrons into and out of a single Au nanoparticle.

\begin{figure}
  \includegraphics[width=8.5cm,keepaspectratio=true]{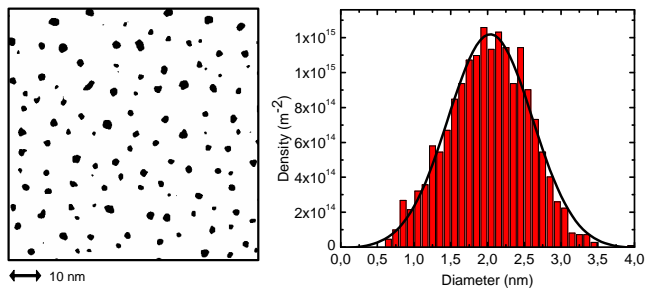}%nl7916bin80nm
\caption{Left: Binarized plane view of self-assembled Au
nanoparticles observed by TEM. See Ref.~\cite{Carrey01-1} for
further information on the binarization process. Right: Size
distribution of the same self-assembly. Fitting gaussian (bold
line), parameters are 2~nm mean diameter and 0.54~nm standard
deviation.} \label{fig:TEMplane}
\end{figure}

The whole structure is elaborated in a sputtering system (base
pressure $5\times 10^{-8}$ mbar) with Ar gas at a dynamic pressure
of $2.5 \times 10 ^{-3}$ mbar. The deposition of a bilayer of
Co(15nm)/Al(0.6nm) is followed by the oxidization of the Al layer in
pure O$_2$ (50 mbar for 10~min) to form the first tunnel barrier.
Then, an ultrathin layer of Au (0.2~nm nominal thickness) is
deposited on top of the bilayer. The 3D growth (see
\cite{Carrey01-1}) of the sputtered gold on top of alumina produces
a self-formed nanoparticles layer. A plane view transmission
electron microscopy (TEM) picture of the Au nanoparticles 2D
self-assembly is shown in Figure~\ref{fig:TEMplane}. The size
distribution of the Au nanoparticles is characterized by a 2~nm
diameter mean value and a 0.5~nm standard deviation with a density
of 1.7~10$^{12}$~cm$^{-2}$. Finally, the Au clusters are capped by
another Al layer (0.6nm) subsequently oxidized in pure O$_2$ with
the same process used to form the first tunnel barrier. This creates
a Co/Al$_2$O$_3$ bilayer with Au nanoclusters embedded in the thin
alumina layer.

To define an electrical contact on a single cluster we use a 4 steps
process combining AFM and optical lithographic techniques (for
further technical details see Ref~\cite{Bouzehouane03}). For the
first step, a photoresist layer of 35~nm is spin coated over the
whole $Co/Al_2O_3/Au/Al_2O_3$ structure. Then, contact zones are
defined using a second photoresist layer and a standard UV
lithography process. During the second step a conductive boron doped
diamond AFM tip is used to nanoindent the thin resist. The
conductance between the conductive sample and the AFM tip is
monitored in real time. In the late stage of the nanoindentation a
tunneling current is established between the tip and the sample. The
exponential thickness dependence allows to precisely control the end
of the indentation process. The tip is then retracted and the sample
is left with a nanometer scale hole on the surface. After this
nano-indentation process, the holes can be inspected by tapping mode
AFM using ultrasharp tips. On figure~\ref{fig:dessin-structure}, we
show the cross section of a typical hole after enlargement by a 30~s
O$_2$ plasma etch. On the cross section one can see that the contact
area section is in the 10~nm range. The samples presented in this
letter use a shorter enlarging plasma etch of 20~s giving holes
sections below the 10~nm range. However, due to a tip nominal radius
of 5~nm, holes having a contact cross-section below 10~nm can not be
properly imaged\cite{Bouzehouane03}. From the nanoparticles density
of 1.7~10$^{12}$.cm$^{-2}$ and assuming a disc shape area for the
contact surface, there is an average of 0.3-1.2 nanoparticles per
nanocontact hole for holes diameter in the 5-10~nm range. The next
step, is the filling of the hole by a sputtered Co 15~nm/ Au~50nm
counter electrode, just after a short O$_2$ plasma. Finally we use
standard optical lithography and ion beam etching techniques to
define the top electrode. This allows us to obtain a \emph{single}
cluster per nanocontact with a high expectation value.

\begin{figure}
  \includegraphics[width=6cm]{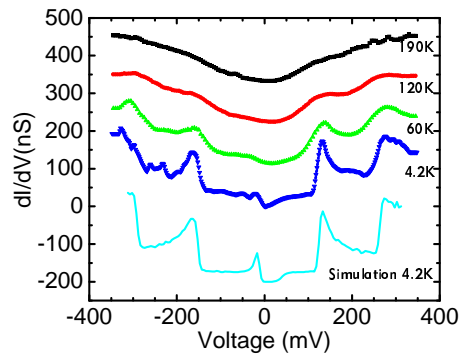}\\
\caption{Differential conductance curves at, 190~K, 120~K, 60~K and
4.2~K for a Co/Al$_2$O$_3$/Au 0.2~nm nominal/Al$_2$O$_3$/Co/Au
nanocontact. Monte-Carlo simulation (plain line) at 4.2~K and with a
background charge offset Q$_0$/e=0.4. The other simulation
parameters are C$_1$=0.4~aF, C$_2$=1.14~aF, R$_2$=3R$_1$.}
  \label{fig:descente-temp}
\end{figure}

The patterned samples were measured in a variable temperature
cryostat from 4.2~K to 300~K. On Figure~\ref{fig:descente-temp}, one
can see the dI/dV(V) curves of a nanopatterned sample at different
temperatures. Above 190~K the conductance curve presents tunnel like
features without any evidence of Coulomb blockade. At 120~K,
shoulders characteristic of Coulomb blockade start to appear in the
conductance curve and get more pronounced as temperature is lowered.
Considering a single nanoparticle, for Coulomb blockade to be
observed at 120~K, the charging energy $E_C$ of the nanoparticle has
to be greater than about $kT_{120K}$ which suggest that the charging
energy is at least $\simeq 10~meV$. Below 60~K an asymmetry in the
shoulder position, typical of the presence of a non zero $Q_0$
background offset charge\cite{Hanna91}, can be observed in the
conductance curves. At 4.2~K, the Coulomb blockade peaks are clearly
visible, with an equal spacing of V$_s\simeq $140~mV between peaks.
As the highest capacitance governs the peak spacing, we extract
C$_{max}$=1.14~aF from V$_s$ through C$_{max}$=e/V$_s$. This
parameter is used as an input for the single nanoparticle
Monte-Carlo simulation (MOSES) \footnote{Monte-Carlo
Single-Electronics Simulator (MOSES), available at
http://hana.physics.sunysb.edu/software/MOSES.htm} which faithfully
reproduces the peaks of the 4.2~K conductance curve in
Figure~\ref{fig:descente-temp}. In this simulation, the structure is
modelized as two tunnel junctions in series. Using
C$_2$=C$_{max}$=1.14~aF and T=$4.2~K$, the best fitting parameters
for the two tunnel junctions are C$_1$=0.4~aF, R$_2$=3R$_1$ and a
background charge of Q$_0/e$=0.4. A V$^2$ term is added to the
simulated curve to take into account the quadratic variation of the
tunnel conductance versus voltage\cite{Brinkman70}. The charging
energy E$_C$ of the nanoparticle can then be calculated assuming the
nanoparticle total capacitance is C$_T$=C$_1$+C$_2$. One finds
$E_C=e^2/2C_T\simeq 50~meV$ which is in agreement with the fading of
the Coulomb blockade features above 120~K. Using electrostatic
simulations and the capacitances extracted from the fit, we can
determine the nanoparticle diameter to be $\simeq 2.5~nm$ in
excellent agreement with the average size determined by TEM.

\begin{figure}
  \includegraphics[width=6cm]{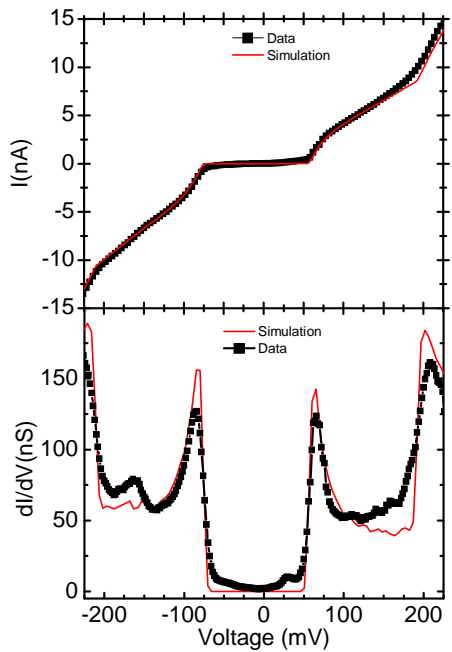}
\caption{Top: I(V) curve(-$\blacksquare$-) measured at 4.2~K for the
sample shown in figure~\ref{fig:descente-temp} after a change in the
background charge. The line is a simulation using the parameters
given in the figure~\ref{fig:descente-temp} caption but with
Q$_0$=0.07~e. Bottom: dI/dV(V) curve (-$\blacksquare$-) obtained
from the derivative of the above I(V) curve. The line represents the
differentiation of the simulation.}
  \label{fig:p1332e4K}
\end{figure}

In Figure~\ref{fig:p1332e4K}, I(V) and dI/dV curves at 4.2~K of the
same sample are shown after a sweep at higher voltage. A change in
the background charge has occurred. The curves are almost symmetric
in voltage bias which reflects the very low offset of the new
background charge. A simulation curve using the same set of
parameters as before, except for a smaller charge offset of
Q$_0$/e=0.07, is also shown on the figure. A clear evidence that we
observe single electron tunneling through a single isolated
nanoparticle is that the same set of junctions parameters are used
to obtain an excellent fit of our data with two different background
charges.

\begin{figure}
  \includegraphics[width=6cm]{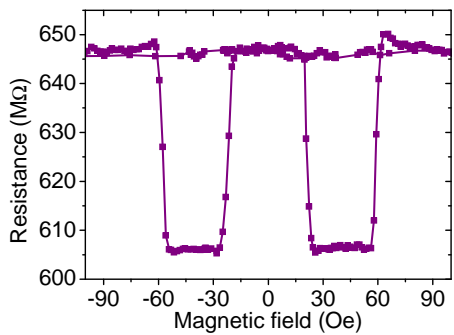}
\caption{Resistance versus magnetic field obtained at 20~mV and 4~K
for the sample of figures~\ref{fig:descente-temp} and
\ref{fig:p1332e4K}.}
  \label{fig:TMR}
\end{figure}

We now focus on spin dependent transport in the sample. In
figure~\ref{fig:TMR}, we present a R(H) curve showing evidence of
tunnel magnetoresistance (TMR). Due to shape anisotropy, the top
"cone" and the bottom plane magnetic electrodes have distinct
coercive fields. Therefore it enables the switching from parallel to
anti-parallel configurations for the electrodes magnetizations. The
occurrence of TMR is a direct proof of spin dependent transport in
the nanostructure, which indicates spin injection from one electrode
into the cluster and spin detection by the second electrode. As the
nanoparticle is \emph{non magnetic},the observation of TMR means
that spin information is conserved in the transport process through
the nanoparticle. The sign of the TMR effect, which rules out direct
tunneling between the two magnetic electrodes, can be understood in
the framework of spin accumulation on the non-magnetic nanocluster
in the co-tunneling regime as discussed in
references\cite{Martinek02,Weymann05}.

In summary, we have developed an original process to investigate the
spin transport properties of a single nanoparticle and provided
evidence for its successful realisation. Our approach paves the way
for a more in-depth study of magneto-Coulomb phenomena in nanosized
clusters.
\begin{acknowledgments}
We acknowledge the support of the EU through the Nanotemplates
project (NMP-CT-2004-505955).
\end{acknowledgments}

%\bibliographystyle{aip}
%\bibliography{apl}% Produces the bibliography via BibTeX.

\end{document}